\documentclass[twocolumn]{jpsj2} 

\title{Pressure-tuning of the $c$-$f$ hybridization in 
Yb metal detected by infrared spectroscopy up to 18~GPa} 

\author{%
Hidekazu \textsc{Okamura}\thanks{E-mail: okamura@kobe-u.ac.jp}, 
Kazuyoshi  \textsc{Senoo}, 
Masaharu \textsc{Matsunami}$^1$ and 
Takao \textsc{Nanba}
}

\inst{%
Graduate School of Science and Technology, Kobe University, 
Kobe 657-8501. 
\\
$^1$RIKEN/SPring-8, Sayo 679-5198.
}

\recdate{\today}

\abst{%
It has been known that the elemental Yb, a divalent metal at 
ambient pressure, becomes a mixed-valent metal under external 
pressure, with its valence reaching $\sim$ 2.6 at 30~GPa.      
In this work, infrared spectroscopy has been used to probe 
the evolution of microscopic electronic states associated 
with the valence crossover in Yb at external pressures up 
to 18~GPa. 
The measured infrared reflectivity spectrum [$R(\omega)$] of 
Yb has shown large variations with pressure.     
In particular, $R(\omega)$ develops a deep minimum in 
the mid-infrared, which shifts to lower energy with 
increasing pressure.    
The dip is attributed to optical absorption due to a 
conduction ($c$)-$f$ electron hybridization state, 
similarly to those previously observed for heavy 
fermion compounds.   
The red shift of the dip indicates that the $c$-$f$ 
hybridization decreases with pressure, which is 
consistent with the increase of valence.     
}

\kword{
Yb metal, valence crossover, infrared synchrotron radiation, 
high pressure, diamond anvil cell
}

\begin{document}
\maketitle

Physical properties of strongly correlated ``heavy fermion'' 
compounds, most typically Ce- or Yb-based compounds containing 
partly filled 4$f$ shell, have attracted much attention.\cite{hewson}     
In heavy fermion compounds, the hybridization between the 
conduction ($c$) electrons and the otherwise localized $f$ 
electrons leads to many interesting phenomena, such as a crossover 
between itinerant and localized characters of the $f$ electrons, 
and the formation of heavy fermion.

Elemental Yb is a divalent metal at ambient pressure, hence 
the $f$ electron configuration is 4$f^{14}$ with no local 
magnetic moment.    However, it has been shown 
using X-ray absorption spectroscopy (XAS) that the valence 
of Yb increases under external pressure, reaching 
$\sim$ 2.6 at 30~GPa.\cite{XAS-1,XAS-2}    
Namely, Yb is a mixed-valent metal in this pressure range.   
Since the number of localized 4$f$ holes increases with 
this valence crossover, it can be viewed from the 4$f$ hole 
point of view as a crossover from an itinerant to a localized 
regime.     Equivalently, it can be also viewed that 
the $c$-$f$ hybridization becomes weaker as the applied 
pressure increases.    
Hence Yb can be regarded as an $f$ electron system 
where an itinerant-localized crossover can be 
caused, and where the $c$-$f$ hybridization can be tuned, 
by the external pressure.      In addition, Yb has 
another interesting property: its resistivity ($\rho$) 
{\it increases} under pressure up to $\sim$ 4~GPa.\cite{Yb-rho}      
It has been suggested by band structure calculation that 
the increase of $\rho$ with pressure is due to a pseudogap 
formation at the Fermi level ($E_F$).\cite{band-calc}

In this work, we have probed the interesting electronic 
structures of Yb metal under pressure using infrared (IR) 
reflectivity [$R(\omega)$] measurement.    IR spectroscopy 
has been quite 
successful in probing the electronic structures associated 
with the $c$-$f$ hybridized state in mixed-valent heavy 
fermion compounds.\cite{basov,degiorgi,schle,okamura}     
We have used a diamond anvil cell (DAC) to produce high pressure.    
It is technically challenging to do reflectivity experiment 
within the limited sample space in a DAC using long wavelength 
IR radiation.     To overcome this, 
we have used an IR synchrotron radiation source and 
an IR microscope at the beam line BL43IR of SPring-8.     
With this combination, a nearly diffraction-limited beam 
diameter ($\sim$ 15~$\mu$m in the mid-IR region) could be 
obtained on the sample.      
The measurements were made in the range 0.03-1.1~eV 
(240-9000~cm$^{-1}$) at room temperature.    
The Yb sample used was a film deposited directly on the diamond 
anvil by evaporating 99.99~\% purity Yb in a high vacuum.   
A gold film deposited next to the Yb film on the diamond anvil 
was used as the reference of $R(\omega)$.    
The films were approximately 50~$\times$ 100~$\mu$m$^2$ wide, 
and 0.5~$\mu$m thick.   
Diamond anvils with a culet size of 0.6~mm were used.  
The pressure was monitored using ruby fluorescence, and 
fluorinert was used as the pressure medium.    
The gasket was a tension-annealed SUS 301 plate with a 
thickness of 0.26~mm.   The pressure variation at different 
portions in the sample space was roughly 1~GPa when 
the average pressure was 18~GPa.   
This variation did not cause a serious problem in our 
analyses since the valence crossover in Yb with pressure 
is known to be very gradual.\cite{XAS-1,XAS-2}

Figures~1(a) and 1(b) summarize the $R(\omega)$ spectra 
taken at different pressures between 0 and 18~GPa.   
\begin{figure}
\begin{center}
\includegraphics[width=0.35\textwidth]{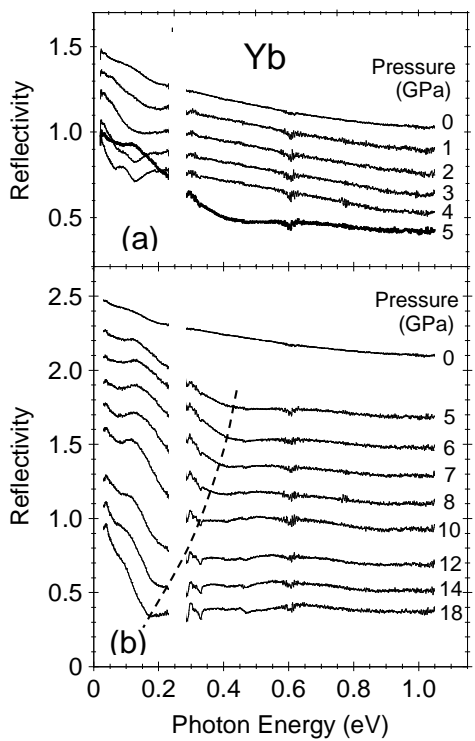}
\caption{
Reflectivity spectra of Yb metal at room temperature 
at external pressures of (a) 0-5~GPa and (b) 0-18~GPa.  
For clarity, each spectrum is offset along the vertical 
axis by 0.1 in (a) and by 0.2 in (b).     The broken curve 
in (b) is a guide to the eye, indicating the shift 
of the dip discussed in the text.    The spectra in the 
0.24-0.28~eV range are absent due to strong 
absorption by diamond, and the weak dip at 0.6~eV is 
an instrumental artifact.   
}
\end{center}
\end{figure}
Before analyzing these spectra, one should note that 
the measurements were done on Yb films {\it directly 
deposited on diamond}, so that the $R(\omega)$ spectra 
in Fig.~1 are 
relative to diamond, in contrast to the usual case of 
$R(\omega)$ relative to vacuum or air.    Due to the large 
refractive index of diamond, $\sim$~2.4, the $R(\omega)$ spectra 
in Fig.~1 are lower than those relative to vacuum.\cite{wooten}     
Hence one should carefully analyze the spectra in 
Fig.~1, since the optical properties of metals are usually 
represented by those relative to vacuum.\cite{wooten}   
Nevertheless, it is quite clear from Fig.~1 that the 
$R(\omega)$ spectra show significant changes with varying 
pressure.   This means that the electronic 
structures of Yb near $E_F$ strongly varies with pressure.    
In this paper we only make qualitative analyses based on 
the $R(\omega)$ in Fig.~1, and more quantitative analyses 
including the correction for diamond's refractive index 
will be presented in a future publication.

Figure~1(a) shows the $R(\omega)$ spectra at pressures 
up to 5~GPa.     In this pressure range, the Yb valence 
determined by XAS does not change very much,\cite{XAS-2} 
but the resistivity of Yb increases with increasing pressure 
until an fcc-bcc structural phase transition occurs at 
approximately 4~GPa.\cite{Yb-rho}     
After the transition, the resistivity drops to a value 
close to that at ambient pressure.\cite{Yb-rho}     
The most significant spectral change in Fig.~1(a) is the 
progressive development of a deep minimum in $R(\omega)$ 
at 0.1-0.2~eV range with increasing pressure, and the sudden 
disappearance of this minimum between 4 and 5~GPa.    Since 
the changes in $R(\omega)$ show good correspondence with those in 
the resistivity, we attribute the changes in $R(\omega)$ to a 
development of pseudogap at $E_F$ with increasing pressure, 
and to the disappearance of the pseudogap upon the structural 
phase transition at $\sim$ 4~GPa.      The magnitude of the 
pressure-induced pseudogap in Yb has been predicted to be 
$\sim$ 0.1~eV by the band calculation.\cite{band-calc}     
This magnitude agrees well with the energy scale of 
the dip in $R(\omega)$.

Figure~1(b) shows the evolution of $R(\omega)$ above 5~GPa.   
The main spectral feature in this pressure range is the 
appearance and shift of another dip in $R(\omega)$, indicated 
by the broken curve: This dip 
appears at $\sim$ 0.45~eV at 5~GPa, and it progressively 
shifts toward lower energy, to below 0.2~eV at 18~GPa.     
A similar dip has been observed in $R(\omega)$ of many 
mixed-valent, heavy fermion 
compounds.\cite{basov,degiorgi,schle,okamura}     
Corresponding to this dip, the mixed-valent compounds have 
a marked absorption peak in their optical conductivity 
$\sigma(\omega)$.\cite{basov,degiorgi,schle,okamura}    
This absorption has been interpreted as resulting from 
optical excitations across the $c$-$f$ hybridization 
gap\cite{hewson} formed in the electronic dispersion 
near $E_F$.\cite{basov,degiorgi,schle,okamura}    
The peak energy of absorption in $\sigma(\omega)$ roughly 
gives the magnitude of the $c$-$f$ hybridization gap, 
and also the hybridization energy.     Since Yb is also 
an mixed-valent metal under high pressure,\cite{XAS-1,XAS-2} 
the dip of $R(\omega)$ in Fig.~1(b) are also likely to result 
from a $c$-$f$ hybridized state, similarly to the case of 
heavy fermion compounds.   This interpretation is consistent 
with the red shift of the dip with increasing pressure, 
since an increase of the Yb valence should be accompanied 
by a decrease of the $c$-$f$ hybridization 
energy.\cite{hewson}   
For more quantitative analyses, however, we need to take 
into account the diamond's refractive index as mentioned 
above, then need to evaluate the optical conductivity.   
This will be done in a future work.

In conclusion, measured infrared $R(\omega)$ of Yb metal 
have revealed two remarkable spectral evolutions under 
pressure: the development of a dip at $\sim$ 
0.15~eV at pressures up to 4~GPa, and that of another dip 
appearing at $\sim$ 0.4~eV at 5~GPa and shifting toward 
lower energy with increasing pressure.     
The former is attributed to a pseudogap formation at $E_F$, 
which is also responsible for the previously reported 
increase of resistivity.   The latter is interpreted in 
terms of a $c$-$f$ hybridized electronic state.   
The shift of the latter dip is taken 
as evidence for a decrease of $c$-$f$ hybridization 
with pressure.

We would like to thank Dr. H. Yamawaki of AIST for providing 
the gasket material.   
The experiments at SPring-8 were done under the approval 
by JASRI (2005B0526 and 2006A1346).    
We acknowledge technical support by the staff at BL43IR.

\end{document}